\documentclass[aps,prd,12pt,amssymb]{revtex4}

\begin{document}

\baselineskip=0.60cm

\newcommand{\ini}{\begin{equation}}
\newcommand{\fin}{\end{equation}}
\newcommand{\inir}{\begin{eqnarray}}
\newcommand{\finr}{\end{eqnarray}}

\def\ol{\overline}
\def\pa{\partial}
\def\ra{\rightarrow}
\def\ts{\times}
\def\df{\dotfill}
\def\bs{\backslash}
\def\dg{\dagger}
\def\la{\lambda}

$~$

\vspace{2 cm}

\title{Seesaw mechanism and leptogenesis 
\footnote{Based on the talk presented at the Laboratoire de
Physique Theorique, Universit\'e de Paris-Sud 11, Orsay Cedex,
France (November 21, 2006).}}

\author{D. Falcone}

\affiliation{Dipartimento di Scienze Fisiche,
Universit\`a di Napoli, Via Cintia, Napoli, Italy}

\begin{abstract}
\vspace{1cm}
\noindent
A brief overview of the phenomenology related to the seesaw mechanism and
the baryogenesis via leptogenesis is presented. In particular, it is explained
how large but not maximal lepton mixing can be achieved within the type II
seesaw mechanism. Moreover, the consequences for leptogenesis are explored,
including flavor effects.
\end{abstract}

\maketitle

\newpage

A breakthrough in particle physics occurred in year 1998, when the SuperKamiokande
Collaboration announced evidence for oscillation of atmospheric neutrinos
$\nu_{\mu}$ \cite{sk}. Then, in 2002, at the Sudbury Neutrino Observatory, evidence
for oscillation of solar neutrinos $\nu_{e}$ has been also found \cite{sno}.
These two important results come after a long series of experiments,
back to the atmospheric neutrino anomaly \cite{ana} and the solar neutrino 
problem \cite{snp}.
The fully terrestrial experiments K2K and KamLAND confirm the above results
for $\nu_{\mu}$ and  $\nu_{e}$, respectively \cite{kk}.

$~$

The most natural explanation of neutrino oscillations \cite{pont} is that
neutrinos have masses and leptons mix just (almost) like quarks do.
In this case, neutrino mass eigenstates $\nu_i$ are related to neutrino
flavor eigenstates $\nu_{\alpha}$ by the unitary transformation
$\nu_{\alpha}=U_{\alpha i}\nu_i$, where U is the lepton mixing matrix
\cite{mns}.

$~$

Oscillation experiments provide square mass differences and mixing angles.
One has
\ini
m_3^2-m_2^2=\Delta m^2_{atm} \simeq (0.05 ~\text{eV})^2
\fin
\ini
m_2^2-m_1^2=\Delta m^2_{sol} \simeq (0.009 ~\text{eV})^2
\fin
and the mixing matrix is given by
\ini
U \simeq
\left( \begin{array}{ccc}
\frac{2}{\sqrt6} &  \frac{1}{\sqrt3} & \epsilon \\
-\frac{1}{\sqrt6} &  \frac{1}{\sqrt3} &  \frac{1}{\sqrt2} \\
\frac{1}{\sqrt6} &  -\frac{1}{\sqrt3} &  \frac{1}{\sqrt2} 
\end{array} \right),
\fin
with $\epsilon < 0.2$, where
$U_{e 2} \simeq \frac{1}{\sqrt3}$ is related to solar oscillations and 
$U_{\mu 3} \simeq \frac{1}{\sqrt2}$ to atmospheric oscillations.

$~$

Moreover, from cosmology (Large Scale Structure and Cosmic Microwave Background)
we get
\ini
\sum m_i \lesssim 0.7 ~\text{eV},
\fin
from the endpoint of tritium beta decay
\ini
U_{e i}^2 m_i^2 \lesssim (2 ~\text{eV})^2,
\fin
and from the neutrinoless double beta decay
\ini
U_{e i}^2 m_i \lesssim 1 ~\text{eV}.
\fin

$~$

Since $m_3^2-m_2^2 \gg m_2^2-m_1^2$, even if the hierarchy is not so strong,
we have three main spectra for the effective neutrinos \cite{alt},
namely the normal
\ini
m_1 < m_2 \ll m_3,
\fin
the inverse
\ini
m_1 \simeq m_2 \gg m_3,
\fin
and the quasi-degenerate
\ini
m_1 \simeq m_2 \simeq m_3.
\fin

$~$

From the summary of experimental informations we note that the neutrino
mass is very small with respect to quarks and charged leptons. Moreover,
about the three independent mixings, $U_{\mu 3}$, $U_{e 2}$, and $U_{e 3}$,
we have $U_{\mu 3}$ maximal, $U_{e 2}$ large but not maximal, and $U_{e 3}$
small. This is in contrast to the quark sector, where the three independent
mixings are small or very small \cite{wol}.

$~$

Both features, small neutrino masses and large lepton mixings, can be accounted
for by the seesaw mechanism \cite{ss}. It requires only a modest extension of
the minimal standard model, namely the addition of a heavy fermionic singlet,
the right-handed neutrino $\nu_R$. In fact, this state allows to create a 
Dirac mass $M_D$ for the neutrino, which is related to the electroweak 
symmetry breaking and thus expected to be of the same order of quark or 
charged lepton masses. Moreover, it allows also a Majorana mass $M_R$ for 
the right-handed neutrino, which is not related to electroweak breaking
(but breaks lepton number) and thus can be very large.
Then the effective neutrino mass matrix is given by
\ini
M_{\nu} \simeq M_D M_R^{-1} M_D=v^2 Y_D M_R^{-1} Y_D,
\fin
where $v$ is the v.e.v. of the standard Higgs field (electroweak breaking
scale), and $Y_D$ is the Yukawa coupling matrix. This seesaw mechanism
is called type I and of course gives small $M_{\nu}$, since $M_D$ is
suppressed by $M_D M_R^{-1}$.

$~$

There is also a triplet seesaw mechanism \cite{mw}, which requires a heavy scalar
triplet $T$. In this case
\ini
M_{\nu}= M_L=Y_L v_L,
\fin
where $v_L$ is an induced v.e.v.,
\ini
v_L=\gamma \frac{v^2}{m_T}.
\fin
The sum of the two terms gives the type II seesaw formula \cite{ss2}
\ini
M_{\nu} \simeq M_D M_R^{-1} M_D +M_L.
\fin
Note that in the type I term there is a double matrix product, which can generate
large mixings from small mixings in $Y_D$ and also in $M_R$ \cite{smir}.
Instead, in the type II (triplet) term, there is only one matrix, so that
large mixing is present or not, by hand.

Then we consider the type I term as fundamental and the triplet term as a kind
of pertubation.

$~$

Let us describe the effect on the mixing in the type II seesaw mechanism
\cite{df1}, using a model for mass matrices \cite{df2}, based on broken $U(2)$
horizontal symmetry and simple quark-lepton symmetry $M_e \sim M_d$,
$M_D \sim M_u$,
\ini
M_D \sim
\left( \begin{array}{ccc}
\la^{12} & \la^6 & \la^{10} \\
\la^6 & \la^4 & \la^4 \\
\la^{10} & \la^4 & 1
\end{array} \right)~m_t,
\fin
\ini
M_e \sim
\left( \begin{array}{ccc}
\la^{6} & \la^3 & \la^{5} \\
\la^3 & \la^2 & \la^2 \\
\la^{5} & \la^2 & 1
\end{array} \right)~m_b,
\fin
\ini
M_R \sim
\left( \begin{array}{ccc}
\la^{12} & \la^{10} & \la^6 \\
\la^{10} & \la^8 & \la^4 \\
\la^6 & \la^4 & 1
\end{array} \right)~m_R,
\fin
with $\lambda=0.2$. Notice that the scale $m_R=M_3 \sim 10^{16}$,
$M_2 \sim 10^{10}$, $M_1 \sim 10^{7}$ GeV.
One could also adopt different mass matrices, keeping the analysis and possibly
the results quite similar.
The type I seesaw mechanism gives
\ini
M_{\nu}^I \sim
\left( \begin{array}{ccc}
\la^4 & \la^2 & \la^2 \\
\la^2 & 1 & 1 \\
\la^2 & 1 & 1
\end{array} \right)~\frac{m_t^2}{m_R},
\fin
corresponding to a normal hierarchy. Moreover we assume that the triplet term is
\ini
M_{\nu}^{II}=M_L=\frac{m_L}{m_R}M_R.
\fin
This form can be motivated within left-right and $SO(10)$ models. However,
in the mood of Ref.\cite{df2} we can think it to be generated by coupling
$M_L$ to the same flavon fields as $M_R$.

$~$

Now, since we do not write coefficients in mass matrices, consider the 
following form of matrix (17),
\ini
M_{\nu}^I \simeq
\left( \begin{array}{ccc}
\la^4 & \la^2 & \la^2 \\
\la^2 & 1+\frac{\la^n}{2} & 1-\frac{\la^n}{2} \\
\la^2 & 1-\frac{\la^n}{2}  & 1+\frac{\la^n}{2}
\end{array} \right)~\frac{m_t^2}{m_R},
\fin
with $n=1,2,3,4$, which gives maximal $U_{\mu 3}$ but different $U_{e 2}$, according
to the value of $n$. In fact, for $n=4$ one has  $\sin \theta_{12}=1/\sqrt2$ and
$\epsilon=0$, that is the bimaximal mixing. For $n=3,2,1$ we get
$\sin \theta_{12} \simeq 0.68; 0.58; 0.25;$ respectively.

$~$

The contribution from $M_{\nu}^{II}$ will we parametrized by the ratio
\ini
k=\frac{m_{\nu}^{II}}{m_{\nu}^{I}} =\frac{m_L m_R}{v^2}=\gamma ~\frac{m_R}{m_T}
\fin
and leads to decrease the mixing 2-3 and especially 1-2 \cite{rod}.

$~$

We consider numerical results, but first the contribution
from $M_e$ should be taken into account. It is similar to the CKM matrix
in the Wolfenstein form and gives
\ini
\sin \theta_{e2} \simeq
\sin \theta_{12} -\frac{\la}{2},
\fin
\ini
\sin \theta_{\mu 3} \simeq \frac{1}{\sqrt2} \left( 1-\la^2 \right),
\fin
\ini
\sin \theta_{e3} \simeq -\frac{\la}{\sqrt2}.
\fin
However, we should also study the Georgi-Jarlskog (GJ) option \cite{gj},
which yields better values for the fermion masses
\ini
M_e \sim
\left( \begin{array}{ccc}
\la^{6} & \la^3 & \la^{5} \\
\la^3 & -3 \la^2 & \la^2 \\
\la^{5} & \la^2 & 1
\end{array} \right)~m_b.
\fin
In this case we get
\ini
\sin \theta_{e2} \simeq
\sin \theta_{12}^{\nu} +\frac{\la}{6},
\fin
$$
\sin \theta_{23} \simeq \frac{1}{\sqrt2} \left( 1-\la^2 \right),
$$
\ini
\sin \theta_{13} \simeq \frac{\la}{3 \sqrt2}.
\fin
Finally we sum $M_{\nu}^{II}$ to $M_{\nu}^{I}$ and combine with the mixing
coming from $M_e$, and look for agreement with experimental ranges of mixings.

We find the following results:

Case $n=4$ requires $0 \le k<0.05$, or $0.08<k<0.18$ for the GJ option.

Case $n=3$ requires $0 \le k<0.04$, or $0.06<k<0.16$ for the GJ option.

Case $n=2$ is reliable only for the GJ choice with $0 \le k<0.10$.

Case $n=1$ is not reliable at all.

Note that the presence of zero means that the triplet term can be absent.
It is instead necessary for $n=4,3$ in the GJ option.

In particular, the difference between the observed quark-lepton complementarity
and the theoretical prediction based on realistic (GJ) quark-lepton symmetry
in cases $n=4,3$ could be ascribed to the triplet contribution within the type II
seesaw mechanism \cite{df3}. We predict
\ini
\theta_{e3} \simeq \lambda / 3 \sqrt2 \simeq 0.05,
\fin
which can be checked in future experiments.

\newpage

Now we consider a cosmological consequence of the seesaw mechanism,
namely the baryogenesis via leptogenesis \cite{fy,oth}. It aims to reproduce
the baryon asymmetry
\ini
\eta_B=\frac{n_B}{n_{\gamma}}= (6.1 \pm 0.2) \cdot 10^{-10}
\fin
by means of the out-of-equilibrium decays of the right-handed neutrinos
which generate a lepton asymmetry partially converted to baryon asymmetry
by electroweak sphaleron processes \cite{krs}.

$~$

The main formulas are
\ini
y_B \simeq \frac{1}{7} \eta_B \simeq \frac{1}{2}y_L
\fin
where
\ini
y_L \simeq 0.3 ~\frac{\epsilon_1}{g_*}
\left( \frac{0.55 \cdot 10^{-3} eV}{\tilde{m}_1} \right)^{1.16}
\fin
in the strong washout regime, or
\ini
y_L \simeq 0.3 ~\frac{\epsilon_1}{g_*}
\left( \frac{\tilde{m}_1}{3.3 \cdot 10^{-3}eV} \right)
\fin
in the weak washout regime, $g_* \simeq 100$,
\ini
\tilde{m}_1 =\frac{(Y_D^{\dg}Y_D)_{11} v^2}{M_1}
\fin
and
\ini
\epsilon_1 \simeq \frac{3}{16 \pi} 
\frac{(Y_D^{\dg}Y_D)^2_{12}}{(Y_D^{\dg}Y_D)_{11}}
\frac{M_1}{M_2}
\fin
is a CP asymmetry related to the decay of the lightest right-handed 
neutrino in the basis where both $M_e$ and $M_R$ are diagonal.

$~$

We have
\ini
Y_D \sim
\left( \begin{array}{ccc}
\la^{7} & \la^5 & \la^{5} \\
\la^6 & \la^4 & \la^2 \\
\la^{6} & \la^4 & 1
\end{array} \right), ~
Y_D^{\dg} Y_D \sim
\left( \begin{array}{ccc}
\la^{12} & \la^{10} & \la^6 \\
\la^{10} & \la^8 & \la^4 \\
\la^6 & \la^4 & 1
\end{array} \right).
\fin
Then, for the Type I matrix model considered above we get
\ini
\epsilon_1=\frac{3}{16 \pi} \lambda^{12} =2.4 \cdot 10^{-10}
\fin
and
$\tilde{m}_1 \simeq m_1 \sim 10^{-4}$ eV, in the weak wash out regime,
so that
\ini
y_B \sim 10^{-14}
\fin
well below the experimental value.

$~$

The contribution from triplet leptogenesis is related to 
$M_1/m_T$ and $k$ \cite{hase}, and hence negligible in our context.

$~$

The formulas used above are valid in the so-called one flavor
approximation. Let us consider also flavored leptogenesis
\cite{bcst}: the processes which wash out lepton number are flavor
dependent (for example, the inverse decays from electrons can destroy
the lepton asymmetry carried by electrons).
Asymmetries in each flavor are washed out differently, and appear with 
different weights in the final formula \cite{ppr}.
Indeed a flavor index $\alpha$ is 
present:
\ini
y_{\alpha \alpha} \simeq 0.3 ~\frac{\epsilon_{\alpha \alpha}}{g_*}
\left( \frac{0.55 \cdot 10^{-3} eV}{\tilde{m}_{\alpha \alpha}} 
\right)^{1.16}
\fin
in the strong washout regime,
\ini
y_{\alpha \alpha} \simeq 0.3 ~\frac{\epsilon_{\alpha \alpha}}{g_*}
\left( \frac{\tilde{m}_1}{3.3 \cdot 10^{-3}eV} \right)
\left( \frac{\tilde{m}_{\alpha \alpha}}{3.3 \cdot 10^{-3}eV} \right)
\fin
in the weak washout regime, with
\ini
\tilde{m}_{\alpha \alpha}=(Y_D^*)_{\alpha 1} (Y_D)_{\alpha 1}
\frac{v^2}{M_1},
\fin
\ini
\epsilon_{\alpha \alpha} \simeq \frac{3}{16 \pi}
\frac{(Y_D^*)_{\alpha 2}(Y_D^{\dg}Y_D)_{12}(Y_D)_{\alpha 1}}
{(Y_D^{\dg}Y_D)_{11}}
\frac{M_1}{M_2}.
\fin
Note that $Y_D$ appears not only in the combination $(Y_D^{\dg}Y_D)$
but also by alone. The final formula for the baryon asymmetry is
\cite{abada}
\ini
y_B=\frac{12}{37} \left( \frac{40}{13} y_{ee}
+ \frac{51}{13} y_{\mu \mu} + \frac{51}{13} y_{\tau \tau} \right).
\fin

$~$

For the type I model considered above we get
$y_{\tau \tau} \simeq y_{\mu \mu} \simeq y_L/2$ and
$y_{ee}$ negligible, with
$\tilde{m}_{\tau \tau} \sim \tilde{m}_{\mu \mu} \sim \tilde{m}_1$
in the weak washout regime, so that $y_B$ is enhanced only by a
factor about 5.

$~$

We see that quark-lepton symmetry seems to imply an amount of baryon
asymmetry which is too small by about three orders of magnitude
\cite{ft,no}. Indeed, $M_1$ is smaller than the Davidson-Ibarra
bound, $M_1 > 10^9$ GeV \cite{dib}.
There are some ways out of this problem. A trivial one is
to consider a moderate hierarchy in $M_D$, for example
$M_D \sim M_e (m_t/m_b)$, see Ref.\cite{df5}. Another is to take
nearly off-diagonal $M_R$ as it appears by inverting the
seesaw formula \cite{bft,afs}. In this case the overall mass
scale is an intermediate scale instead of a unification scale.

$~$

Another different possibility is to take into account special flavor 
effects (allignement conditions) which preserve $\epsilon_2$, contrary
to usual assumptions \cite{sfe}. Then for our model we get
\ini
\epsilon_2 \simeq \frac{3}{16 \pi} \lambda^8 =1.5 \cdot 10^{-7}
\fin
and $y_B \sim 10^{-11}$ in the right order of magnitude.

$~$

In conclusion, the (type II) seesaw mechanism is able to explain small
neutrino masses and large or maximal lepton mixings.
However, if one assumes quark-lepton symmetry, the amount of baryon
asymmetry produced in the leptogenesis mechanism is too small,
unless special flavor effects are realized, or $M_R$ has a nearly
off-diagonal form.

$~$

$~$

We thank Luis Oliver and Asmaa Abada in Orsay for kindly hospitality and
discussions, and Franco Buccella in Naples for discussions and support.

\end{document}